\documentclass[aps,twocolumn,prb,floatfix]{revtex4-1}
\usepackage[a4paper,top=2.0cm,bottom=2.00cm,left=2.0cm,right=2.0cm]{geometry}
\usepackage{amsmath}
\usepackage{amsfonts}
\usepackage{amssymb}
\usepackage{esint}
\usepackage{mathtools}
\usepackage{bbold}
\usepackage{graphicx}
\usepackage{color}
\usepackage{mathptmx}
\usepackage{multirow}
\usepackage{tikz}
\usetikzlibrary{arrows.meta}
\def\kappatnsr{\tikz{\draw[{<[length=.8mm,width=.9mm]}-{>[length=.8mm,width=.9mm]}] (0,0) -- (2.1mm,0) node[midway, below,inner sep=0, yshift=-1.5pt, xshift=-0.2pt] {$\kappa$};}}

\newenvironment{talign*}
 {\let\displaystyle\textstyle\csname align*\endcsname}
 {\endalign}

\newcommand{\nn}{\nonumber}

\def\ns{\negthickspace}

\newcommand{\grad}{\mbox{\boldmath$\nabla$}}

\def\t3{{\widehat{\tau}_3}}
\def\t2{{\widehat{\tau}_2}}
\def\t1{{\widehat{\tau}_1}}
\def\tz{{\widehat{\tau}_3}}
\def\ty{{\widehat{\tau}_2}}
\def\tx{{\widehat{\tau}_1}}
\def\tone{\widehat{1}}

\renewcommand{\Re}{\mbox{Re}}
\renewcommand{\Im}{\mbox{Im}}
\def\Tr#1{\mbox{Tr}\left\{#1\right\}}
\def\ns{\negthickspace}
\def\superscript{\scriptsize}
\def\whg{\widehat g}
\def\whga{{\widehat g}^{\mbox{\small a}}}
\def\whgK{{\widehat g}^{\mbox{\superscript K}}}
\def\whgRA{{\widehat g}^{\mbox{\superscript R,A}}}
\def\whgR{{\widehat g}^{\mbox{\superscript R}}}
\def\whgA{{\widehat g}^{\mbox{\superscript A}}}
\def\gRA{g^{\mbox{\superscript R,A}}}
\def\gR{g^{\mbox{\superscript R}}}

\def\fRA{f^{\mbox{\superscript R,A}}}

\def\LambdaRA{\Lambda^{\mbox{\superscript R,A}}}
\def\SigmaRA{\Sigma^{\mbox{\superscript R,A}}}

\def\whtR{\widehat{t}^{\mbox{\superscript R}}}
\def\whtA{\widehat{t}^{\mbox{\superscript A}}}

\def\whx{\widehat{x}}

\def\whxR{\widehat{x}^{\mbox{\superscript R}}}
\def\whxA{\widehat{x}^{\mbox{\superscript A}}}
\def\whxK{\widehat{x}^{\mbox{\superscript K}}}
\def\whxa{\widehat{x}^{\mbox{\small a}}}
\def\whSigma{\widehat\Sigma}
\def\whSigmaRA{\widehat\Sigma^{\mbox{\superscript R,A}}}

\def\whSigmaa{\widehat\Sigma^{\mbox{\small a}}}
\def\DRA{D^{\mbox{\superscript R,A}}}

\def\DeltaRA{{\tilde{\Delta}^{\mbox{\superscript R,A}}}}
\def\eRA{{\tilde{\varepsilon}^{\mbox{\superscript R,A}}}}
\def\e{\varepsilon}
\def\vell{\bm{\ell}}
\def\p{\mathbf{\hat p}}

\def\vp{\mathbf{p}}

\def\vr{\mathbf{r}}

\def\vv{\mathbf{v}}

\def\He{$^{3}$He}
\def\Hea{$^{3}$He-A}
\def\upt{UPt$_{3}$}
\def\sro{Sr$_{2}$RuO$_{4}$}

\def\imp{\text{imp}}
\def\point#1#2{{\tt #1}_{\mbox{\tiny #2}}}

\def\ber{\begin{eqnarray}}
\def\eer{\end{eqnarray}}
\def\be{\begin{equation}}
\def\ee{\end{equation}}

\def\nn{\nonumber}
\def\bmat{\begin{pmatrix}}
\def\emat{\end{pmatrix}}
\def\bsmat{\left(\begin{smallmatrix}}
\def\esmat{\end{smallmatrix}\right)}

\def\kb{k_{\mbox{\tiny B}}}
\usepackage{hyperref}
\DeclareGraphicsExtensions{.eps, .pdf, .jpg, .tif}
\usepackage{float}
\begin{document}
\title{Impurity-Induced Anomalous Thermal Hall Effect in Chiral Superconductors}
\author{Vudtiwat~Ngampruetikorn} 
\email{current address: Initiative for the Theoretical Sciences, 
                        The Graduate Center, CUNY, New York, NY 10016;
       email: vngampruetikorn@gc.cuny.edu} 
\author{J.~A. Sauls}
\email{email: sauls@northwestern.edu}
\affiliation{Center for Applied Physics \& Superconducting Technologies, 
             Department of Physics, Northwestern University, Evanston, IL 60208 
             and \\
             Fermi National Accelerator Laboratory, Batavia, IL 60510}
\date{\today}

%------------------------------------------------------------------------------
\begin{abstract}
Chiral superconductors exhibit novel transport properties that depend on the topology of the order parameter, topology of the Fermi surface, the spectrum of bulk and edge Fermionic excitations, and the structure of the impurity potential. 
In the case of electronic heat transport, impurities induce an anomalous (zero-field) thermal Hall conductivity that is easily orders of magnitude larger than the quantized edge contribution.
The effect originates from branch-conversion scattering of Bogoliubov quasiparticles by the chiral order parameter, induced by potential scattering. The former transfers angular momentum between the condensate and the excitations that transport heat.
The anomalous thermal Hall conductivity is shown to depend to the structure of the electron-impurity potential, as well as the winding number, $\nu$, of the chiral order parameter, $\Delta(p)=|\Delta(p)|\,e^{i\nu\phi_{\p}}$. 
The results provide quantitative formulae for interpreting heat transport experiments seeking to identify broken T and P symmetries, as well as the topology of the order parameter for chiral superconductors.
\end{abstract}
\maketitle
\allowdisplaybreaks
%------------------------------------------------------------------------------

{\it Introduction --}
Chiral superconductivity occurs when bound pairs of Fermions condense into a macroscopically occupied two-particle state, $\psi(\vr)\sim(x+iy)^{\nu}\sim e^{i\nu\phi}$, corresponding to Cooper pairs circulating about a unique chiral axis, $\vell$, with angular momentum $\nu\hbar$. Mirror symmetry (P), with respect to a plane containing the chiral axis $\vell$, is spontaneously broken in combination with time-reversal (T) symmetry. Thus, left- and right-handed Cooper pairs form time-reversed ground-states with counter-circulating currents.\cite{vollhardt90}
Superfluid $^3$He-A is currently the only BCS condensate that is firmly established to exhibit chiral pairing. The identification of broken P and T symmetries was made by the observation of anomalous Hall transport of electrons moving through \Hea,\cite{ike13} in quantitative agreement with transport theory.\cite{she16}

The search for an electronic analog of the chiral phase of superfluid \He\ has been widely pursued,\cite{mac03,mae12,str09} driven in part by theoretical predictions of novel properties of topological superconductors.
In 2D materials, chiral d-wave superconductivity is predicted for doped graphene,\cite{nan12,bla15} while a chiral p-wave state is proposed for MoS$_2$.\cite{yua16}
For the 3D pnictide, SrPtAs, there is evidence of broken T symmetry from $\mu$SR;\cite{bis13} a chiral d-wave state has been proposed theoretically.\cite{fis14}
The perovskite superconductor, \sro, is a promising candidate for chiral superconductivity based on $\mu$SR and Kerr rotation measurements.\cite{luk98,xia06} 
The first superconductor reported to show evidence of broken T symmetry was the heavy fermion superconductor, \upt, based on $\mu$SR linewidth measurements.\cite{luk93} This experiment followed theoretical predictions of broken T and P symmetries in the B-phase of \upt, i.e. the lower temperature superconducting phase.\cite{hes89}
Particularly striking is the observation of the onset of Kerr rotation at the transition to the low-temperature B-phase of \upt.\cite{sch14}
However, definitive proof of \emph{bulk} chiral superconductivity in any of these materials awaits a zero-field bulk transport measurement that otherwise vanishes in the absence of broken P and T symmetries.  

Chiral superconductors are also topological phases characterized by a Chern number equal to the winding number, $\nu$, of the phase of the Cooper pairs. 
The non-trivial topology manifests as $|\nu|$ branches of chiral Fermions confined near a boundary at which the topology changes discontinuously.\cite{vol10,miz16}
These edge states are unique to chiral superconductors.
The response of the edge spectrum to a thermal gradient has been shown to generate an \emph{anomalous} (zero field) thermal Hall conductance, $K_{xy}^{\mbox{\tiny edge}}=\nu\frac{\pi}{6}\,\kb^2\,T/\hbar$, in which the chiral axis $\vell$ assumes the role of the perpendicular magnetic field.\cite{rea00,nom12,sum13,gos15}
While the zero-energy edge state is protected by the bulk topology, the spectrum of chiral edge states, and their transport currents, is sensitive to surface disorder.
Furthermore, impurities embedded in an otherwise fully gapped chiral superconductor can destroy the bulk topology by closing the bulk gap. When this happens the thermal Hall conductance persists, but is no longer quantized.
Theoretical work based on point-like impurities predicts an anomalous thermal Hall effect (ATHE) in chiral p-wave superconductors, but no ATHE for $|\nu| \ge 2$.\cite{arf88,li15b,yip16}

In this Letter we present a theory of anomalous Hall transport of heat in chiral superconductors with impurity disorder, and show that the impurity-induced ATHE can be orders of magnitude larger than that from the chiral edge states. 
We also show the impurity-induced ATHE requires the coupling between quasiparticles, which transport heat and charge, with the condensate which breaks T and P symmetries. 
Two mechanisms provide the coupling between quasiparticles and the chiral condensate. 
The first is the transfer of Cooper pair angular momentum to quasiparticle transport currents via branch-conversion (Andreev) scattering.
When an incident electron ($e$) with angular momentum $m\hbar$ relative to an impurity undergoes branch conversion scattering the outgoing hole ($h$) acquires angular momentum $\nu\hbar$ from the chiral condensate, i.e., $e_m \rightarrow h_{m+\nu}$. Similarly, $h_m\rightarrow e_{m-\nu}$. 
This process requires finite scattering cross sections for partial waves associated with the angular momenta of incident and outgoing states and is therefore absent for point-like impurities which generate only s-wave scattering. 
The second mechanism is the direct coupling of the perturbation to the condensate, which is possible if, and only if, the perturbation and the condensate belong to the same orbital representation. A thermal gradient generates a p-wave perturbation, $\propto\mathbf{v}_{\mathbf{p}}\cdot\nabla T$, and will couple directly to a chiral p-wave condensate. 
Importantly, both mechanisms must also breaks particle-hole symmetry to allow a net transfer of angular momentum between the condensate and scattered quasiparticles.
As a result, for point-like impurities, the absence of Andreev scattering means that an ATHE is possible only for chiral superconductors with $|\nu|=1$. 
However, for finite-radius impurities, scattering in multiple angular momentum channels leads to an ATHE for chiral superconductors with larger Chern numbers, $|\nu|\ge 2$. 
This is also the basic mechanism responsible for the anomalous Hall effect of electrons moving through superfluid \Hea.\cite{she16}

%-----------------------------------------------------------------------------------

{\it Theory --}
Our theory and analysis starts from the quasiclassical formulation of the transport equations for nonequilibrium superconductivity,\cite{eil68,lar75} with our notation and formalism explained in Ref.~\onlinecite{nga19b}.
We calculate the thermal conductivity tensor for chiral superconductors from the non-equilibrium response of the quasiparticle distribution and spectral functions to a thermal gradient in the linear-response 
limit, in which case the heat current is $\mathbf{j}_\e=-\kappatnsr\cdot\grad T$ where $\kappatnsr$ is the thermal conductivity.

The effects of impurity scattering on the chiral ground state, the appearance of a sub-gap quasiparticle spectrum, and the non-equilibrium response to a temperature gradient, are encoded in our theory via the impurity averaging technique, and the resulting quasiparticle-impurity $t$-matrix. The latter is a functional of the quasiclassical propagator and self energies, all of which are calculated self-consistently, including the impurity-scattering vertex corrections.
To highlight the effects of chirality on heat transport, we focus on fully gapped 2D chiral superconducting ground states defined on a cylindrically symmetric Fermi surface, $\Delta(\p)=\Delta\,e^{i\nu\phi_{\p}}$, where $\phi_{\p}$ is the azimuthal angle of relative momentum, $\vp$, of the Cooper pair and $\nu$ is the winding number of the order parameter around the Fermi surface.\footnote{We assume the normal state has P and T symmetry, and consider spin-singlet states, $\hat{\Delta}(\p)=(i\sigma_y)\Delta(\p)$, and ``unitary'' spin-triplet states of the form, $\hat{\Delta}(\p)=(i\vec{\sigma}\sigma_y\cdot\hat{d})\,\Delta(\p)$, where ${\Delta}(\p)$ has even (odd) parity for singlet (triplet) pairing.}
The mean-field Hamiltonian for excitations in a chiral ground state takes the form
\begin{equation}
\widehat H=\xi_\vp\tz + \Delta (\tx\cos\nu\phi_{\p}+\ty\sin\nu\phi_{\p}),
\end{equation}
where $\xi_\vp$ is the normal-state dispersion and $\tx,\ty,\tz$ denote the Pauli matrices in particle-hole space.\footnote{We removed the spin matrices by a unitary transformation. See Ref.~\onlinecite{nga19b}.}  
The spectra of quasiparticles and Cooper pairs are also encoded in the equilibrium retarded 
(R) and advanced (A) propagators, 
\begin{equation}\label{eq:g_eq}
\whgRA_\text{eq}(\p;\e)
=-\pi
\frac{\eRA\tz-\DeltaRA e^{i\tz\nu\phi_{\p}}(i\ty)}
     {\sqrt{(\DeltaRA)^2-(\eRA)^2}}
\,,
\end{equation}
which define the corresponding quasiparticle and Cooper pair propagators, $\gRA$ and $\fRA$, as $\whgRA=-\pi[\gRA(\e)\tz+\fRA(\e)e^{i\tz\nu\phi_{\p}}(i\ty)]$.
Note that $\eRA$ and $\DeltaRA$ are the renormalized excitation energy and order parameter, $\eRA=\e\pm i0^{+}-\SigmaRA$ and ${\DeltaRA}=\Delta+\LambdaRA$.
We consider only renormalization by impurity scattering since this is the dominant scattering process in the low-temperature limit. The structure of the impurity self energy in Nambu space takes the form,\footnote{The unit Nambu matrix self-energy {$\DRA(\varepsilon)\tone$} drops out of the equilibrium propagator but contributes to the a.c. linear response of the superconductor.}
\begin{equation}
\hspace*{-3.05mm}
\whSigmaRA_{\imp}(\p;\e) 
\ns\equiv\ns 
\DRA(\e)\tone\ns +\ns \SigmaRA(\e)\tz\ns+\ns\LambdaRA(\e)e^{i\tz\nu\phi_{\p}}(i\ty).
\end{equation}

The mean-field order parameter, $\Delta$, satisfies the weak-coupling gap equation, $\Delta=-\frac{V}{2}\fint d\e\tanh\frac{\e}{2T}\Im f^R(\e)$, where the integration is over the pairing bandwidth, $(-\varepsilon_c,+\varepsilon_c)$, and $V$ is the strength of the pairing interaction, $V(\p,\p')=2V\cos[\nu(\phi_{\p}-\phi_{\p'})]$ for the $\nu^{th}$ irreducible representation of $\point{SO(2)}{\ns}$ symmetry of the Fermi surface.

For a homogeneous, random distribution of impurities the self-energy, $\widehat\Sigma_{\imp}(\p;\e)=n_{\imp}\widehat{t}(\p,\p;\e)$, is proportional to the mean impurity density, $n_{\imp}$, and the forward-scattering limit of the the single-impurity $t$-matrix, the latter of which satisfies,
\begin{eqnarray}
\hspace*{-5mm}
\widehat t(\p',\p)
&=&
\widehat t_N(\p',\p) 
\nonumber\\
&+& 
N_f
\textstyle\ns\displaystyle\int_0^{2\pi}\ns\frac{d\phi_{\p''}}{2\pi}\,
\widehat t_N(\p',\p'')\left[\whg(\p'')-\whg_N\right]\widehat t(\p'',\p)
\,.
\ns\ns
\label{eq:tmat_eq}
\end{eqnarray}
We omit the superscripts unless needed, $N_f$ denotes the single-spin normal-state density of states at the Fermi energy and $\whgRA_N =\mp i\pi\tz$ is the normal-state propagator. The normal-state $t$-matrix is parametrized in terms of quasiparticle-impurity scattering phase shifts, $\delta_m$, for each angular momentum channel, $m$,
\begin{equation}
\widehat{t}_N(\p',\p)=\frac{-1}{\pi N_f}
\sum_{m=-\infty}^{+\infty}\frac{e^{im(\phi_{\p}-\phi_{\p'})}}{\cot\delta_{m}+\whg_N/\pi}
\,.
\end{equation}
For the equilibrium propagator (Eq.~\ref{eq:g_eq}), we obtain the $t$-matrix from \eqref{eq:tmat_eq}, and the corresponding impurity self-energy terms,
\begin{eqnarray}
\Sigma(\e)  &=& 
\sum_m\mathcal{A}_m\,g(\e)\sin^2\delta_m
\\
\Lambda(\e) &=& 
\sum_m\mathcal{A}_m\,f(\e)\sin\delta_m\cos(\delta_{m}-\delta_{m+\nu})\sin\delta_{m+\nu}
\\
D(\e) &=&
\sum_m\mathcal{A}_m\sin\delta_m
[
g(\e)^2\sin\delta_m\sin(\delta_m-\delta_{m+\nu})
\nn\\[-.5ex]
&\ns&
\qquad\qquad\qquad\quad
+\cos\delta_m\cos(\delta_m-\delta_{m+\nu})
],
\\
\mathcal{A}_m&=&
-\frac{n_\imp}{\pi N_f}
 \frac{\tilde\Delta^2-\tilde\e^2}{\tilde\Delta^2\cos^2(\delta_{m}-\delta_{m+\nu})-\tilde\e^2}
\,.
\end{eqnarray}
For a single impurity, multiple scattering results in sub-gap quasiparticle bound states, $\e_{b,m}=\pm|\Delta|\cos(\delta_{m}-\delta_{m+\nu})$, which appear as isolated poles of the $t$-matrix amplitude $\mathcal{A}_m(\varepsilon)$. These states broaden into sub-gap bands for finite impurity density.
The off-diagonal self-energy, $\Lambda$, is generated by Andreev scattering, a branch-conversion scattering process in which a particle turns into a hole, or vice versa. The angular momentum associated with each partial wave of the incoming and outgoing states must differ by an integer equal to the Cooper pair angular momentum quantum number $\nu$, i.e., \emph{both} $\delta_m$ and $\delta_{m+\nu}$ must be finite for a given value of $m$.
Thus, point-like impurities, which scatter only in the $s$-wave channel, do not generate branch-conversion processes in chiral superconductors, and so do not couple to Cooper pair angular momentum.

Heat current in response to an imposed temperature gradient is obtained from the non-equilibrium response of the Keldysh propagator, $\delta\whg^K$,
\begin{equation}
\label{eq:j_e}
\mathbf{j}_\e=N_f 
\int_0^{2\pi}\frac{d\phi_{\p}}{2\pi}
\int\frac{d\e}{4\pi i}\,
\e\vv_{\vp}\,\Tr{\delta\whgK(\p;\e)}
\,,
\end{equation}
where $\vv_{\vp}=v_f\p$ is the Fermi velocity. It is convenient to introduce the \emph{anomalous} propagator, $\whga$, and anomalous self-energy, $\whSigmaa$, defined in terms of the corresponding Keldysh (K), retarded (R) and advanced (A) functions, $\delta\whxa = \delta\whxK-\tanh\frac{\e}{2T}\,(\delta\whxR-\delta\whxA)$, where $\whx\in\{\whg,\whSigma\}$. The first-order corrections to the retarded and advanced propagators and self-energies vanish to linear order in $\vv_{\vp}\cdot\grad\Phi$ (cf.~Ref.~\onlinecite{gra96a}). Thus, the linear response contribution to the anomalous propagator reduces to the Keldysh propagator,
\begin{eqnarray}
\delta\whga=-\frac{C^{a}_+\whgR_\text{eq}/\pi + D^{a}_{-}}{(C^{a}_+)^2+(D^{a}_-)^2}
&\big[&
(\whgR_\text{eq}-\whgA_\text{eq})\,i\hbar\vv_{\vp}\cdot\nabla\Phi
\nn\\[-.5ex]
&+&
(\whgR_\text{eq}\delta\whSigmaa - \delta\whSigmaa\whgA_\text{eq})
\big],
\label{eq:dga}
\end{eqnarray}
where $\nabla\Phi=\nabla\tanh[\e/2T(\vr)]$ is the gradient of the local equilibrium distribution function, $C^a_+=2\Re\sqrt{\tilde\Delta^2-\tilde\e^2}$ and $D^{a}_-=2i\,\Im D(\e+i0^+)$. 

The non-equilibrium response of the self-energy is obtained from the anomalous $t$-matrix, which in linear response reduces to 
\begin{equation}\label{eq:linear_tmatrix}
\hspace*{-2mm}\delta\whSigmaa(\p) \ns=\ns n_{\imp}\,N_f\ns
\int_0^{2\pi}\ns
\frac{d\phi_{\p'}}{2\pi}\,
\whtR_\text{eq}(\p,\p')\delta\whga(\p')\whtA_\text{eq}(\p',\p)
\,.
\end{equation}
These are the ``vertex corrections'' in diagrammatic quantum field theories. They describe the dynamical screening of perturbations by long-wavelength collective excitations.\cite{rai94b} This self energy correction is central to anomalous Hall transport.
In its absence the diagonal terms of the Keldysh propagator have the same angular dependence as the perturbation (cf.~Eq.~\ref{eq:dga}), and thus generate a heat current along the temperature gradient and no Hall response.

%Move to end of intro
%
%The impurity-induced ATHE requires the coupling between quasiparticles, which transport heat and charge, with the condensate which breaks T and P symmetries. This occurs via two mechanisms. The first is the transfer of Cooper pair angular momentum to quasiparticle transport currents via branch-conversion (Andreev) scattering. The second is the direct coupling of the perturbation to the condensate. The latter is possible if, and only if, the perturbation and the condensate belong to the same orbital representation. For a thermal gradient the perturbation, $i\hbar\vv_{\vp}\cdot\grad\Phi$, is p-wave and will couple directly to a chiral p-wave condensate. 
%%
%As a result, for point-like impurities (s-wave scatterers) Andreev scattering is absent, and thus an ATHE is possible only for chiral superconductors with $|\nu|=1$. However, for finite-radius impurities scattering in channels with $|m|\le m_{\text{max}}\approx\mbox{Int}[k_f R]$ leads to an ATHE for chiral superconductors with larger Chern numbers, $|\nu|\ge 2$.

%------------------------------------------------------------------------------------------
\begin{figure}
\centering
\includegraphics[width=\linewidth]{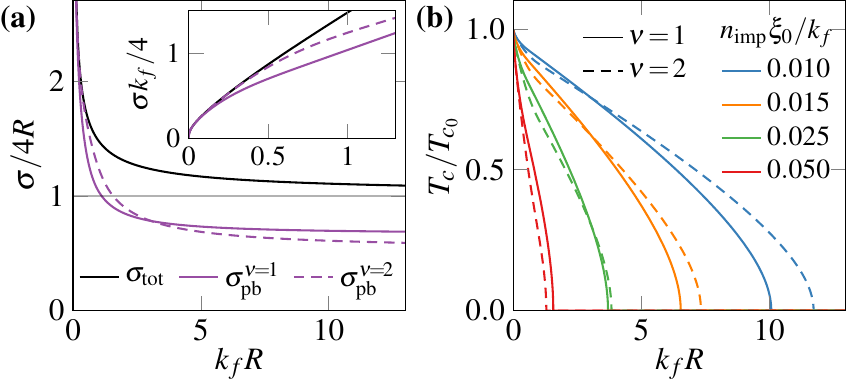}
\caption{
(a) The total and pair-breaking scattering cross sections, $\sigma_\text{tot}$ and $\sigma_\text{pb}$, as functions of the hard-disc radius $R$ for $\nu=1$ (solid) and $\nu=2$ (dashed).
(b) The critical temperature $T_c$ as a function of $k_fR$ for $\nu=1$ (solid) and $\nu=2$ (dashed), and a range of impurity densities (see legend).
}
\label{fig:Tc_harddisc}
\end{figure}
%-----------------------------------------------------------------------------------

{\it Impurity Scattering Model --} To quantify the effects of finite-size impurities, we consider hard-disc scattering characterized by the scattering phase shifts, $\tan\delta_m = J_{|m|}(k_fR)/N_{|m|}(k_fR)$, where $R$ is the hard-disc radius, and $J_m(z)$ ($N_m(z)$) are Bessel functions of the first (second) kind.\cite{lap86}
Non-magnetic impurities in chiral superconductors are pair-breaking.\cite{lar65,thu98} The critical temperature, $T_c$, is suppressed, $\ln\frac{T_{c_0}}{T_c}=\Psi\left(\frac{1}{2}+\frac{1}{2}\frac{\xi_0\sigma_\text{pb}n_\imp}{T_c/T_{c_0}}\right)-\Psi\left(\frac{1}{2}\right)$, where $\Psi(x)$ is the digamma function,\cite{abramowitz70} and $T_{c_0}$ and $\xi_0=\hbar v_f/2\pi T_{c_0}$ are the critical temperature and coherence length in the clean limit. The pair-breaking cross section is given by $\sigma_\text{pb}=(2/k_f)\sum_{m=-\infty}^{+\infty}\,\sin^2(\delta_m-\delta_{m+\nu})$, for a chiral order parameter with a winding number $\nu$. The pair-breaking cross-section vanishes for $s$-wave superconductors ($\nu=0$), yielding $T_c=T_{c_0}$ as expected.\cite{and59} In Fig.~\ref{fig:Tc_harddisc}(a) the pair-breaking cross section is shown to differ substantially from the total cross section, $\sigma_\text{tot}=(4/k_f)\sum_m\sin^2\delta_m$, except in the limit $k_fR\ll 1$. In Fig.~\ref{fig:Tc_harddisc}(b) the dependence of $T_c$ on both the impurity radius and the winding number are highlighted for several impurity densities. 

%------------------------------------------------------------------------------------
\begin{figure}
\centering
\includegraphics[width=\linewidth]{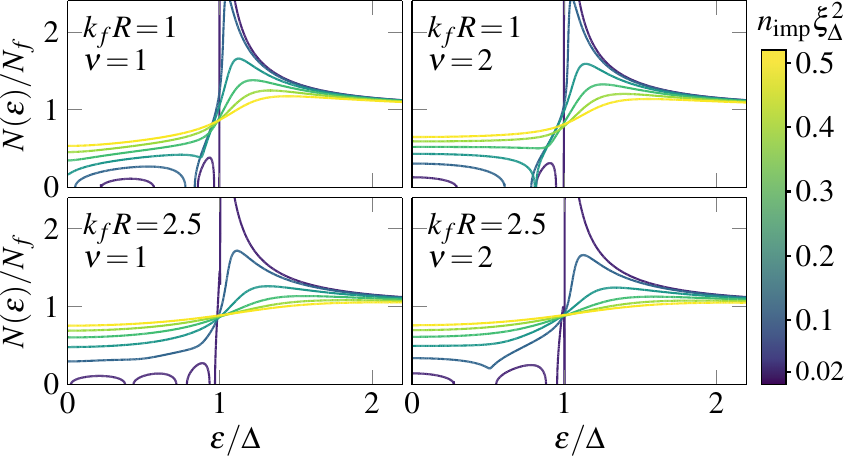}
\caption{
$N(\varepsilon)/N_f$ for chiral states with $\nu=1$ (left) and $\nu=2$ (right), various impurity densities normalized by $\xi_\Delta^2=(\pi N_f\Delta)^{-1}$ (see legend), and impurity radii $k_fR=1$ (top) and $2.5$ (bottom).
}
\label{fig:dos}
\end{figure}
%-----------------------------------------------------------------------------------
%-----------------------------------------------------------------------------------
\begin{figure}
\centering
\includegraphics[width=\linewidth]{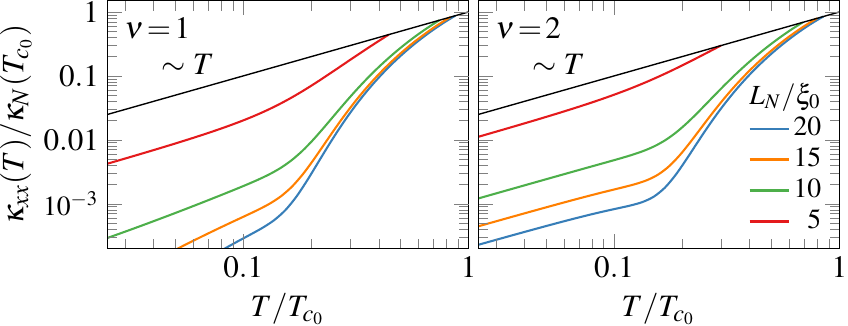}
\caption{
$\kappa_{xx}/\kappa_N(T_{c_0})$ versus $T/T_{c_0}$ for 
$\nu=1$ (left) and $\nu=2$ (right), $k_f R=1$ and various values of $L_N/\xi_0$
(see legend). The normal-state thermal conductivity is shown in black.
\label{fig:kappaxx_log}
}
\end{figure}
%-----------------------------------------------------------------------------------

{\it Density of States --}
The quasiparticle density of states, $N(\e)=N_f\Im\gR(\e+i0^+)$, also depends on the chiral winding number. Figure~\ref{fig:dos} shows sub-gap bound states, broadened into bands by the finite impurity density. 
These states are formed via multiple Andreev scattering by the chiral order parameter, induced by potential scattering. The number of the sub-gap bands, and their bandwidths, depend not only on the structure of the impurity potential, e.g. the hard-disc radius, but also on the chiral winding number. 
This fact has important implications for thermal transport in the limit $T\lesssim\Delta$. Impurities enhance the thermal conductivity of the superconducting state at low temperatures through the formation of sub-gap states that transport heat. A sub-gap ``metallic'' density of states at the Fermi energy, $N(0)\ne 0$, results in $\kappa_{xx}\propto T$. Figure~\ref{fig:kappaxx_log} shows the temperature dependence of $\kappa_{xx}$.
The low-temperature metallic behavior is always present for $\nu=2$, whereas for $\nu=1$ it occurs only at sufficiently high impurity densities.

The normal-state thermal conductivity, $\kappa_N=(\pi^2/3)N_f (v_f T) L_N $, is limited by the transport mean free path, $L_N=1/(\sigma_\text{tr}\,n_{\imp})$, where the transport cross section is defined by $\sigma_\text{tr}=(2/k_f)\sum_m\sin^2(\delta_{m}-\delta_{m+1})$.
In the superconducting state, axial symmetry is broken by the chiral order parameter. The corresponding thermal conductivity tensor, $\kappa_{ij}$, acquires off-diagonal terms, $\kappa_{xy}=-\kappa_{yx}$, in addition to the diagonal components, $\kappa_{xx}=\kappa_{yy}$. Thus, there is a transverse (Hall) component of heat current.
The longitudinal and transverse conductivities, $\kappa_{xx}$ and $\kappa_{xy}$, are obtained by computing the heat current induced by a temperature gradient using Eqs.~\ref{eq:j_e}-\ref{eq:linear_tmatrix}.

\begin{figure}[h]
\centering
\includegraphics[width=\linewidth]{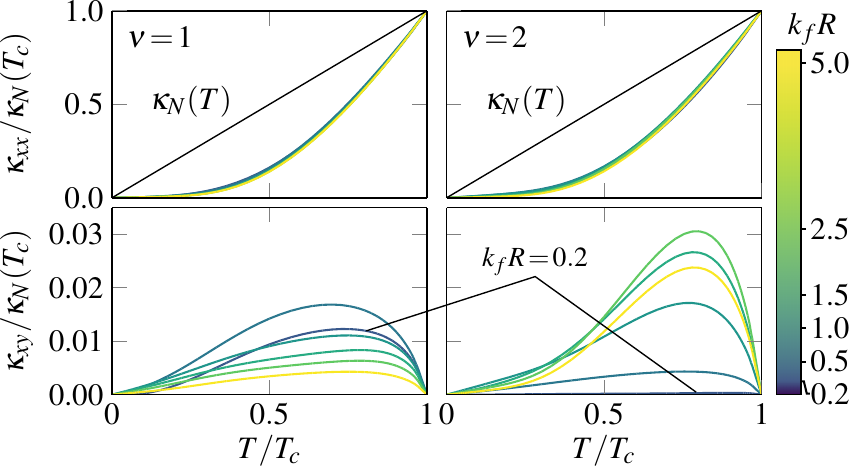}
\caption{Longitudinal (top) and transverse (bottom) thermal conductivity versus $T/T_c$ for $\nu=1$ (left) and $\nu=2$ (right). The normal-state transport mean free path is $L_N/\xi_0=7.5$, and various impurity radii (see legend). The black line is $\kappa_N(T)/\kappa_N(T_c)$.
\label{fig:kappa_xy_Tc}
}
\end{figure}

Figure~\ref{fig:kappa_xy_Tc} shows the effects of finite-size impurities on heat transport. While the longitudinal conductivity, $\kappa_{xx}$, is only weakly affected by impurity size or winding number (except at ultra-low temperatures), the thermal Hall conductivity, $\kappa_{xy}$, depends strongly on both $k_f R$ and $\nu$.
For impurities that are smaller than the inverse Fermi wavelength, $k_f R < 1$, quasiparticle scattering is predominantly in the $s$-wave channel. The resulting thermal Hall conductivity is strongly suppressed for winding number $\nu=2$, but remains finite in the limit $k_f R \rightarrow 0$ for $\nu=1$. The numerical results agree with our previous observation that the ATHE vanishes in the limit of point-like impurities for chiral superconductors with $|\nu|\ge 2$.
For impurities with $k_f R>1$ the Hall conductivity is substantially larger for chiral superconductors with $\nu=2$, compared to $\nu=1$. Also note that the Hall conductivity is sensitive to 
the impurity potential, in this case exhibiting nonmonotonic dependence on the impurity size. 

%-----------------------------------------------------------------------------------
%-----------------------------------------------------------------------------------

%\vspace*{-5mm}

\begin{figure}[h]
\centering
\includegraphics[width=\linewidth]{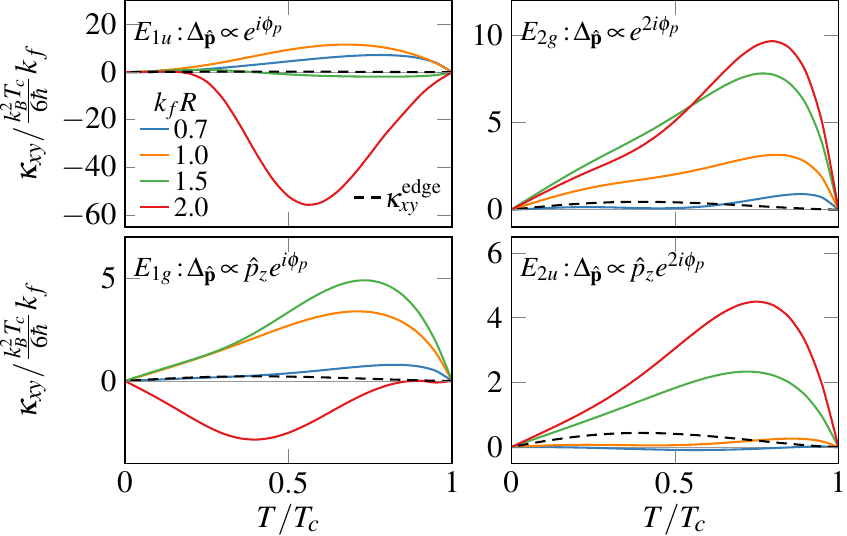}
\caption{
%Thermal Hall conductivity, 
$\kappa_{xy}$ vs $T/T_{c}$ for various $k_fR$ (see legend), and chiral states belonging to the $E_{1u}$, $E_{2g}$, $E_{1g}$ and $E_{2u}$ representations of $D_{6h}$. 
The dashed lines represent the edge contribution to $\kappa_{xy}$. Results are shown for $L_N/\xi_0\!=\!7.5$ and $k_f\xi_0\!=\!100$.
}
\label{fig:Kxy_3d}
\end{figure}

{\it 3D Chiral Superconductors --} The results for 2D chiral states easily generalize to chiral states defined on closed 3D Fermi surfaces with line and point nodes. This includes the ATHEs in 3D candidates for chiral superconductors, including \sro\ and \upt. 
Figure~\ref{fig:Kxy_3d} shows the thermal Hall conductivity for chiral superconductors belonging to the spin-triplet, odd-parity $E_{1u}$ and $E_{2u}$ representations, and the spin-singlet, even-parity $E_{1g}$ and $E_{2g}$ representations of the hexagonal $D_{6h}$ point group, and $E_{u}$ and $E_{g}$ representations of $D_{4h}$. These representations cover nearly all of the proposed candidates for chiral superconductors.
Note that the impurity-induced thermal Hall conductivity (solid lines) typically dominates the edge contribution~\cite{qin11,gos15} (dashed lines) in all chiral pairing states with finite-size impurities.
Also, note the sensitivity of $\kappa_{xy}$ to $k_f R$ particularly for winding number $|\nu|=1$, as well as the order of magnitude difference in $\kappa_{xy}$ for chiral $E_{1u}$ versus $E_{1g}$.
For \upt\ with $k_f\!=\!1\,\text{\AA}^{-1}$, $\xi_0\!=\!100\,\text{\AA}$ and $T_c\!=\!0.5\,\text{K}$ we estimate $\kappa_{xy}\!>\!3\!\times\!10^{-3}\,\text{WK$^{-1}$m$^{-1}$}$ at $T=0.8T_c$ for the $f$-wave $E_{2u}$ chiral state with a hard-sphere impurity radius $k_fR\!=\!1.5$ (Fig.~\ref{fig:Kxy_3d}), which is well within reported sensitivities of current experimental measurements of the thermal Hall effect.\cite{hir15}

{\it Bulk vs.~Edge --}
The impurity-induced ATHE typically dominates the edge contribution for a 2D chiral p-wave superconductor by an order of magnitude or more depending on the impurity density and material parameters. 
For $k_f\xi_0=100$, $L_N/\xi_0=7.5$ and $k_fR=0.5$, we have 
$\kappa_{xy}^\text{imp}\approx100\kappa_{xy}^\text{edge}$ 
at $T=0.8 T_c$ (maximum in $\kappa_{xy}$, see Fig.~\ref{fig:kappa_xy_Tc}).
Here $\kappa_{xy}^\text{edge}$ is computed from Eq.~(25) in Ref.~\onlinecite{qin11}.
However for sufficiently clean 2D chiral p-wave superconductors the edge contribution, given by the quantized value, $\kappa_{xy}^\text{edge}/T=\pi k_B^2/6\hbar$,\cite{rea00,sen99,sum13}
can dominate the impurity contribution at very low temperatures.
In the case of the latter, $\kappa^{\text{imp}}_{xy}/T$ vanishes due to the absence of sub-gap states at $\e=0$. Thus, below a threshold impurity density, the dominant contribution to the ATHE for the fully gapped chiral p-wave case at $T\ll \Delta$ comes from the response of the chiral edge Fermions.

%-----------------------------------------------------------------------------------
%-----------------------------------------------------------------------------------

{\it Conclusions --} 
Branch-conversion (Andreev) scattering by the chiral order parameter is the key mechanism responsible for skew scattering, and thus the thermal Hall conductivity, in chiral superconductors. For finite size impurities Andreev scattering is activated for any winding number, e.g. $\nu=1$ (p-wave) or $\nu=2$ (d-wave). 
The impurity-induced thermal Hall conductivity is easily orders of magnitude larger than that due to edge states.
In summary, our work provides quantitative formulae for interpreting heat transport experiments seeking to identify broken T and P symmetries, as well as the topology of the order parameter for chiral superconductors.

{\it Acknowledgements --} The research of JAS was supported by the National Science Foundation (Grant DMR-1508730). The research of VN was supported through the Center for Applied Physics and Superconducting Technologies. We thank Pallab Goswami for discussions.

%------------------------------------------------------------------------------
%\bibliographystyle{apsrev4-1_PRX_style}
%\bibliography{QFS,CM,QM,Books}
%------------------------------------------------------------------------------
%
\end{document}